\documentclass[conference]{IEEEtran}

\usepackage[letterpaper, left=0.680in, right=0.680in, bottom=1.049in, top=0.71in]{geometry}

\usepackage{ifpdf}
\ifCLASSINFOpdf
\usepackage[pdftex]{graphicx}
\DeclareGraphicsExtensions{.pdf,.jpeg,.png}
\else
\usepackage[dvips]{graphicx}
\DeclareGraphicsExtensions{.pdf}
\usepackage[caption=false, font=footnotesize]{subfig}
\fi

\usepackage{xcolor,soul,framed} 
\usepackage[noadjust]{cite}

\usepackage{xcolor,soul,framed}
\usepackage[noadjust]{cite}

\colorlet{shadecolor}{yellow}

\usepackage[pdftex]{graphicx}
\graphicspath{{../pdf/}{../jpeg/}}
\DeclareGraphicsExtensions{.pdf,.jpeg,.png}

\usepackage{amssymb,mathtools}
\usepackage{amsmath}
\usepackage{array}
\usepackage{mdwmath}
\usepackage{mdwtab}
\usepackage{eqparbox}
\usepackage{url}
\usepackage[ruled,vlined,linesnumbered]{algorithm2e}
\usepackage{enumitem} 
\usepackage[mathcal]{euscript}
\usepackage{tabularx} 
\usepackage{multirow} 
\usepackage{booktabs}
\usepackage{makecell} 
\usepackage{aligned-overset} 

\usepackage{url}
\usepackage{multicol}

\usepackage{balance}
\usepackage{bbm}
\usepackage{amsmath}
\usepackage{amsthm} 

\DeclareMathOperator*{\argmax}{argmax}
\usepackage{pifont}

\graphicspath{{./Figures/}}

\renewcommand{\vec}[1]{\boldsymbol{\mathrm{#1}}}

\begin{document}

\title{DRL-Based Dynamic Channel Access and SCLAR Maximization for Networks Under Jamming}

\author{\IEEEauthorblockN{
             Abdul~Basit\IEEEauthorrefmark{1}, Muddasir Rahim\IEEEauthorrefmark{1}, 
			Georges~Kaddoum\IEEEauthorrefmark{1}$^,$\IEEEauthorrefmark{3}, 
			Tri~Nhu~Do\IEEEauthorrefmark{2}, and Nadir~Adam\IEEEauthorrefmark{1}
			}
		
		\IEEEauthorblockA{
		\IEEEauthorrefmark{1}
		Department of Electrical Engineering, \'{E}cole de Technologie Sup\'{e}rieure (\'{E}TS), Universit\'{e} du Qu\'{e}bec, Montr\'{e}al, QC, Canada, \\Emails: abdul.basit.1@ens.etsmtl.ca, muddasir.rahim.1@ens.etsmtl.ca,
			georges.kaddoum@etsmtl.ca, nadir.adam@etsmtl.ca}
   \IEEEauthorrefmark{3} Artificial Intelligence \& Cyber Systems Research Center, Lebanese American University\\   
  \IEEEauthorrefmark{2} Department of Electrical Engineering, Polytechnique Montreal, QC, Canada, Email: tri-nhu.do@polymtl.ca
	}

\maketitle
\begin{abstract} 
This paper investigates a deep reinforcement learning (DRL)-based approach for managing channel access in wireless networks. Specifically, we consider a scenario in which an intelligent user device (iUD) shares a time-varying uplink wireless channel with several fixed transmission schedule user devices (fUDs) and an unknown-schedule malicious jammer. The iUD aims to harmoniously coexist with the fUDs, avoid the jammer, and adaptively learn an optimal channel access strategy in the face of dynamic channel conditions, to maximize the network's sum cross-layer achievable rate (SCLAR). Through extensive simulations, we demonstrate that when we appropriately define the state space, action space, and rewards within the DRL framework, the iUD can effectively coexist with other UDs and optimize the network's SCLAR. We show that the proposed algorithm outperforms the tabular Q-learning and a fully connected deep neural network approach.  
\end{abstract}

\section{Introduction}
The expansion of wireless technologies and the growing demand for high-speed data are contributing to making modern networks more complex. As networks become increasingly complex, effective network management becomes more crucial for operational efficiency. The medium access control (MAC) layer is pivotal for optimizing data transmission because it effectively manages multiple users' access to a shared channel. In wireless networks, the fair and optimal allocation of limited resources among users is essential, but malicious users like jammers disrupt this process. Jammers, in particular, severely compromise channel utilization and network performance, which impacts throughput, delay, and quality of service (QoS) \cite{khadr2022jamming}. Conventional channel access methods like time-division multiple access (TDMA) and carrier sense multiple access (CSMA) often underperform in heterogeneous networks due to their inability to adapt to changing network conditions.

Prior research has extensively investigated using artificial intelligence (AI)-based solutions to address MAC-layer challenges. For instance, \cite{miuccio2022learning} advocates improving the performance of wireless networks using machine learning (ML)-based solutions to manage radio resources. In \cite{da2020noma}, a Q-learning random-access strategy is introduced for ultra-dense networks with non-orthogonal multiple access (NOMA). It involves a centralized base station (BS) utilizing observed channel status to select optimal transmission power. Similarly, in \cite{mennes2020multi}, deep reinforcement learning (DRL) is employed to predict spectrum usage in cognitive radio (CR) networks to reduce collisions with neighboring networks, thus enhancing network performance.

The authors of \cite{xin2022deep} propose a deep learning-based CSMA protocol for Wi-Fi networks, and the authors of \cite{fihri2020machine} focus on using supervised learning to detect MAC-layer attacks in cognitive radio networks (CRNs) to improve performance in their respective types of networks. The authors of \cite{khairy2020constrained}, on the other hand, put forward a DRL-based unmanned aerial vehicle (UAV)-altitude random access approach for Internet of Things (IoT) networks. Finally, the authors of \cite{yu2019deep} propose to use a DRL-based smart node that can coexist with other nodes utilizing slot-based protocols to ensure adaptive access. Addressing performance degradation involves introducing intelligent user devices (iUDs) that can learn optimal channel access strategies on the fly and enhance network performance by reducing collisions and jamming susceptibility. More specifically, AI and ML techniques enable iUDs to make informed transmission decisions in dynamic wireless settings.

To this end, we propose a DRL-based iUD that can coexist with legitimate fixed-schedule user devices (fUDs) and jammers in a quasi-static network. In the network considered, all UDs, including fUDs and the iUD, transmit to an access point (AP) through a shared time-slotted uplink channel, and the jammer disrupts communication between the UDs and the AP. The uplink data channel undergoes frequent transformations, which causes the transmit power, channel coefficients, and path loss parameters of all UDs and jammers to vary. Consequently, each UD's signal-to-interference-plus-noise ratio (SINR) and achievable rate continually fluctuate. We formulate the channel access problem as a partially observable Markov decision process (POMDP) in a constantly changing channel environment with jammers and use a deep Q-network (DQN) to approximate the optimal action-value function~\cite{morales2020grokking}.

Our proposed approach enables the iUD to effectively coexist with fUDs and avoid jammers, thus enhancing channel utilization and the network's sum cross-layer achievable rate (SCLAR) by minimizing collisions and jamming susceptibility. The methodology involves formulating the channel access problem as a POMDP in dynamic channel conditions with jammers. We employ a residual deep neural network (ResDNN) for optimal action-value function approximation and conduct simulations to demonstrate that our proposed approach outperforms the iUD that uses a tabular Q-learning approach, or a fully connected deep neural network (FC-DNN) to approximate the action-value function in terms of channel utilization, SCLAR, and jamming robustness.  
\section{System Model}\label{section: system model}
\begin{figure}[t]
\centering
 \includegraphics[width=\linewidth]{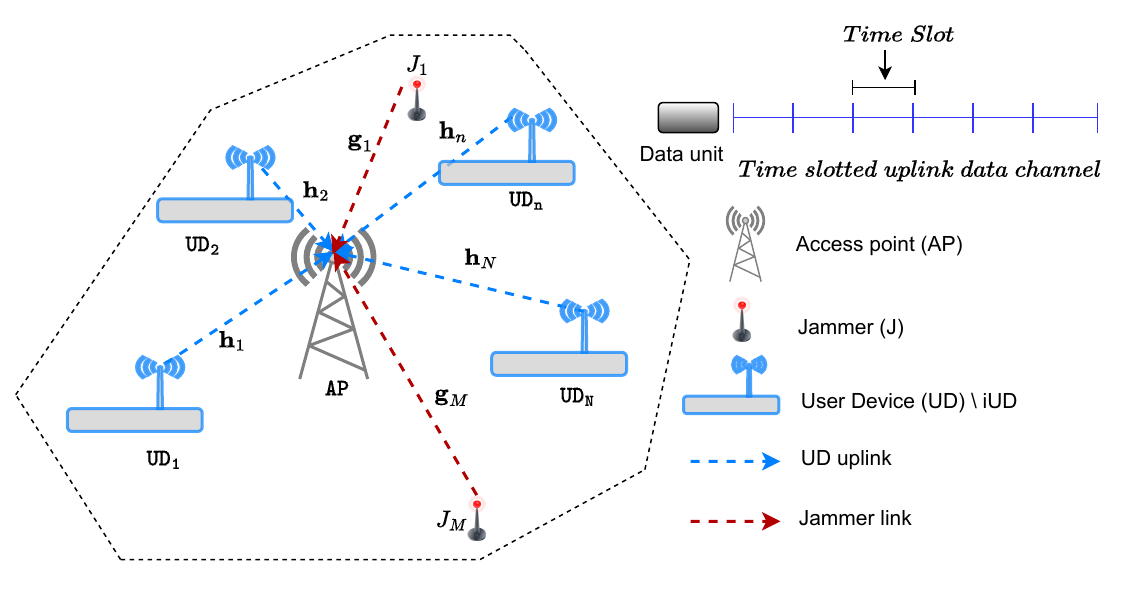}
 \caption{Illustration of the network with legitimate and malicious user devices communicating with the access point.}
 \label{fig:network_model}
 \end{figure}

We consider uplink transmission in a quasi-static wireless network in which each UD's transmission schedule changes slowly over time. There is one $\mathrm{AP}$ with $K$ antennas, \mbox{$N$ single-antenna UDs}, and $M$ jammers in the network in question. The UDs transmit data packets to the $\mathrm{AP}$ via a shared time slotted ($t= \{0, 1, 2, 3, \cdots\}$) uplink channel, and the jammer always tries to attack the data channel, as illustrated in Fig.~\ref{fig:network_model}. Let $[\vec{A}]_{n,n}\in \{0, 1\}$ denote the transmission status of the $n^{th}$ UD, where $1$ means that the UD is transmitting a data packet in time slot $t$ and $0$ means that it is not transmitting any packet. Furthermore, the status of the $m^{th}$ jammer $\mathrm{J}_m$ in time slot $t$ can be denoted by $[\vec{I}_\mathrm{J}]_{m,m}\in \{0, 1\}$ with $1$ meaning the jammer is active and $0$ meaning it is not. At the end of each time slot, the $\mathrm{AP}$ broadcasts an acknowledgment packet (ACK) to the UDs over a separate control channel. It is assumed that the jammer has no impact on the control channel.
\subsection{Slotted Time-Based MAC}
It is assumed that all the UDs are synchronized with the $\mathrm{AP}$. The total time is split into $T= \{T_1, T_2, \cdots, T_f, \cdots, T_F\}$ repeated frames, where the subscript $\{1,2, \cdots, f, \cdots, F\}$ denotes the frame number, and $F$ is the total number of frames. Moreover, a single frame consists of a fixed number of time slots. For instance, the $f^{th}$ frame is divided into $t_f = \{t^1_f, t^2_f, \cdots, t^s_f, \cdots, t^S_f\}$ slots, where the superscript $\{1,2, \cdots, s, \cdots, S\}$ represents the time slot number and $S$ denotes the total number of time slots in the frame, i.e., $t^s_f$ is the $s^{th}$ time slot in the $f^{th}$ frame. 
Packet transmission starts at the beginning of a time slot, and a single packet takes up to a one time slot to transmit. Furthermore, each UD can transmit multiple packets in a single frame. Thus, multiple UDs can share a radio frequency channel at time slot $t$. 
In the event multiple UDs, i.e., $\mathrm{UD}_n$ and $\mathrm{UD}_{n'} \in \mathcal{N} \backslash \{\mathrm{UD}_n\}$ try to transmit simultaneously in the time slot $t$, a collision happens. Additionally, if at a given time slot $t$ $[\vec{I}_{\tt J}]_{m,m}$ = 1 and $\mathrm{UD}_n$ transmits a packet, jamming occurs and $\mathrm{UD}_n$'s packet is lost.
\subsubsection{Stochastic \textnormal{fUDs}}
In the network in question, an fUD is a type of UD that has a predefined transmission schedule that has been assigned in a stochastic manner. Consider a coefficient [$\chi$] that equals 1 if $\mathrm{UD}_n$ transmits a packet in time slot $t^s_f$ and 0 otherwise. This coefficient models the transmission status of $\mathrm{UD}_n$, which follows a Bernoulli distribution that is parameterized by $\Omega$, i.e., $[\chi] \sim $Bern$(\Omega)$. Therefore, the signal to generate an fUD's transmission vector in a frame $t_f= [0, 1, 0, 0, 1, ...]$ can be represented mathematically as
\begin{align} 
P([\vec{A}]_{n,n}) = 
    \begin{cases} 
        1 - \Omega &  [\chi] = 0 \\ \Omega & [\chi] = 1.
    \end{cases} 
\end{align}
\subsubsection{Random Jamming}
 We consider there to be a random jammer present in the network that transmits jamming signals intermittently according to a predetermined attack strategy with the sole intention of disrupting fUDs and iUD's transmission. The jammer transmits in predefined $t_f^{s}$ time slots out of total $T_F$ slots and repeats its jamming pattern from frame to frame. The jammer's jamming schedule $I_{\tt J}$ in the $F^{th}$ frame is denoted as $\vec{I}^F_J$ and can be expressed as 
 \begin{align}
 \vec{I}^{F}_{\tt J} = 
 \begin{cases}
 1, & \big[(n-1) S_o\big] \leq t^s_f \leq \big[(n-1) S_o + S_{o'}\big] \\ 
 0, & \big[(n-1) S_o + S_{o'}\big] \leq t^s_f \leq [n S_o],
 \end{cases}
 \end{align}
 where $ n \in \mathbb{N}$ is the period number, which is set to a predefined value. Furthermore, $S_{o'} \in S, S_{o'} < S_o$, and $S_{o'} = [0, S_{o'}]$ is the time period in which the jammer is inactive and the transmission of $\mathrm{UD}_n$ is thus unaffected. Similarly, $S_{o} \in S$, and $S_{o} = [S_{o}, S]$ is the time period in which the jammer is active and any packets transmitted by $\mathrm{UD}_n$ are destroyed. Furthermore, it holds that $S_{o'} + S_{o} = S$.
\subsubsection{DRL-Based iUD} 
The DRL-based iUD is a UD that uses our proposed channel access policy. The iUD is oblivious to 1) the number of coexisting legitimate UDs, 2) the transmission schedules of coexisting legitimate UDs, 3) the presence and quantity of jammers, and 4) the jammers' operating mechanism. The iUD's goal is to learn the transmission schedule of coexisting legitimate UDs and jammers in order to opportunistically enact an optimal transmission schedule and enhance network performance. 
\subsection{Signal Modelling of the Network Considered }
\subsubsection{Signal Received at the $\mathrm{AP}$}
Let $\vec{h}_{n}\in \mathbb{C}^{K \times 1}$ and $\vec{g}_m\in \mathbb{C}^{K \times 1}$ be the channel vectors between the $n^{th}$ UD and the $\mathrm{AP}$ and the $m^{th}$ jammer and the $\mathrm{AP}$, respectively. Furthermore, the channel matrices between the $N$ UDs and the $\mathrm{AP}$ and the $M$ jammers and the $\mathrm{AP}$ are written as $\vec{H} \in \mathbb{C}^{K \times N}$ and $\vec{G} \in \mathbb{C}^{K \times M}$, respectively. We ignore large-scale fading and assume that the elements of $\vec{H}$ are independent and identically Gaussian distributed with zero mean and unit variance. 
Let $\vec{A}$ denote the network operation-action matrix of $N$ UDs and, as defined earlier, $[\vec{A}]_{n,n}$ is the transmission status of the $n^{th}$ UD.
Let $x_{n}$ be the signal transmitted by the $\mathrm{UD}_n$ to the $\mathrm{AP}$. Furthermore, the transmitted signal vector $\vec{x}_{\mathrm{UD}}=[x_1, \cdots, x_n, \cdots, x_N] \in \mathbb{C}^{N \times 1}$ is an $[N \times 1]$ column vector that contains the transmitted signals. The transmit power matrix $\vec{P}_{\mathtt{UD}} = \operatorname{diag} ( P_1, P_2, \cdots, P_n, \cdots, P_N)$ is an $[N \times N ]$ diagonal matrix that contains the transmit power of each UD in the diagonal elements where ${P}_n$ represents the transmit power of $\mathrm{UD}_n$. Since $N$ UDs share the same time-frequency resources, the received signal vector $\vec{y}$ is a $[K \times 1]$ column vector that contains the signals received at the $\mathrm{AP}$ from all $N$ UDs. Therefore, the signal received at the $\mathrm{AP}$ is given by
\begin{align} \label{eq_rx_signal}
 \vec{y} =\!\! {\sum_{n=1}^{N} [\vec{A}]_{n,n} \sqrt{{P_n}} \vec{h}_{n} x_{n}}+\sum_{m=1}^{M} [\vec{I}_{\tt J}]_{m,m} \sqrt{{P_{m}}} \vec{g}_{m} x_{m}\!\!\!+\!\vec{n},
\end{align}
where $\vec{n} \in \mathbb{R}^{K \times 1}$
denotes the additive white Gaussian noise (AWGN) vector at the $\mathrm{AP}$ with $E[|\vec{n}\vec{n}^T|^2]=\sigma^2\vec{I}_K$,  $\vec{A}= \operatorname{diag} \big[[\vec{A}]_{1,1},[\vec{A}]_{2,2}, \cdots, [\vec{A}]_{{N},{N}}\big]$
 is an $[N \times N]$ diagonal matrix and 
$\vec{I}_{\tt J}= \operatorname{diag} \big[[\vec{I}_{\tt J}]_{1,1},[\vec{I}_{\tt J}]_{2,2}, \cdots, [\vec{I}_{\tt J}]_{M_{\tt J}, M_{\tt J}}\big]$ 
is an $[M_{\tt J} \times M_{\tt J}]$
diagonal matrix. 
\subsubsection{Linear Detection at the $\mathrm{AP}$ and SINR Formulation}
Let $\hat{\vec{x}}_\mathtt{UD} \in \mathbb{C}^{N \times 1}$ be estimated using a linear receiver. If we consider the linear detection matrix $\vec{V}\in \mathbb{C}^{K \times N}$, which is used to separate the signal received at the $\mathrm{AP}$ into $N$ streams, $\hat{\vec{x}}_\mathtt{UD}$ can be expressed as
\begin{align} 
\label{eq_decoding_matrix}
\scalebox{0.94}{$\hat{\vec{x}}_\mathtt{UD} = \vec{V}^{\sf H} \vec{y} =\textstyle \vec{V}^{\sf H}\vec{H} \sqrt{{\vec{P}_\mathtt{UD}}} \vec{A}\vec{x}_{\mathtt{UD}}+\vec{V}^{\sf H}\vec{G} \sqrt{{\vec{P}_{\mathtt{J}}}}\vec{I}_{\mathtt{J}}\vec{x}_{\tt J} + \vec{V}^{\sf H} \vec{n}$}.
\end{align}

If we consider matched filter successive interference cancellation (MF-SIC), the SINR for $\mathrm{UD}_n$, $\Gamma_{n}$, can be obtained as follows by removing the interference from other UDs with MF-SIC.
\begin{align}\label{mf_n}
\scalebox{1}{$
  \Gamma_{n}
 = \frac{[\vec{A}]_{n,n} P_{n} || h_{n}||^4} 
 {
  \sum\limits_{\substack{n'=1\\ n'\neq n }}^{N}\!\! [\vec{A}]_{n',n'} {P_{n'}} |\vec{h}^{\sf H}_{n}\vec{h}_{{n'}}|^2 
\!\!+\!\!\!\sum\limits_{\substack{m=1}}^{M} \!\!\vec{I}_{\tt J_{m}} {P_{m}}|\vec{h}^{\sf H}_{n} \vec{g}_{m}|^2
\!\!+\!\!||\vec{h}_{n}||^2 \delta^2}.
$}
\end{align}

\section{Problem Formulation}\label{section: problem_formulation}

\subsection{Sum Cross-Layer Achievable Rate Formulation}
Cross-layer achievable rate (CLAR) is a technique that jointly considers physical-layer and MAC-layer issues in an integrated framework. The physical-layer achievable rate of $\mathrm{UD}_n$ in a single time slot $t^s_f$ can be represented as
\begin{align}\label{eq: physical_rate_slot}
 C_{_n}[t^s_f] = \log_2 ( 1 + \Gamma_n[t^s_f] ),
\end{align}
where $C_{n}[t^s_f]$ is the achievable rate in bits per slot per hertz. Furthermore, the achievable rate of $\mathtt{UD}_n$ for the frame $t_f$ can be expressed as
\begin{align}\label{eq: physical_rate_frame}
    \vec{C}_{n}[t_f] &= \textstyle\sum_{n=1}^{{N}} \sum_{s=1}^{S} C_{n}[t^s_f].
\end{align}
Successful transmission rate ($\xi$) is a MAC-layer metric that measures the number of data packets that are successfully transmitted by $\mathrm{UD}_n$ to the $\mathrm{AP}$ over a given period of time and can be expressed as
\begin{align}
    \xi_{n}[t_f] =\frac{p_{\mathtt{S}-\mathtt{UD}_n}[t^1_f] + p_{\mathtt{S}-\mathtt{UD}_n}[t^2_f], \cdots, p_{\mathtt{S}-\mathtt{UD}_n}[t^S_f]}{S},
\end{align} 
where $p_{\mathtt{S}-\mathtt{UD}_n}[t^s_f]$ is the probability of successful transmission of $\mathtt{UD}_n \rightarrow \mathtt{C}$ in time slot $t^s_f$. The CLAR of $\mathtt{UD}_n$ in time slot $(t^s_f)$ is denoted by $R_n[t^s_f]$ and can be expressed as
 \begin{align}
 R_{n}[t^s_f]= \xi_{\mathtt{UD}_n}[t^s_f] C_{n}[t^s_f].
\end{align} 
 Furthermore, $\vec{r}_n \in \mathbb{R}^{T\times 1}$ represents the CLAR of an entire frame and can be expressed as
\begin{align}
    \vec{r}_n = [R_n [t^1_f], R_n [t^2_f], \cdots, R_n [t^s_f] ,\cdots , R_n [t^S_f]].
\end{align}
 
 The SCLAR maximization (SCLARM) is an optimization problem that aims to find the optimal resource allocation to maximize the SCLAR which we elaborate on next.
\subsection{SCLARM Problem Formulation}
The network contains $N$ legitimate UDs, i.e., $|\mathcal{N}| = N$, of which $(N-1)$ are fUDs and one is an iUD. The UD action vector $\vec{a}_n \in \{0,1\}^{T \times 1}$, where $\vec{a}_n = \{ [\vec{A}_{n} (t) : k = 1,..., |\mathcal{C}|, n=1,...,N, t = 0,...,T-1 \}$, is the network's UD action vector of a given $\mathrm{UD}_n$ over a given time period $T$ and describes the UD's operation. The SCLAR maximization (SCLARM) optimization problem can be written as
\begin{subequations}\label{opt}
 \begin{alignat}{2}
 &\underset{\vec{A}}{\mathrm{maximize}}
 &\quad 
 & \textstyle\sum_{n=1}^{|\mathcal{N}|} (\vec{r}_n)^{\sf T} \vec{a}_n,  \label{eq_obj_func}\\
 &\text{subject to} 
 && a_{n} (t) \in \{0, 1\}, \forall t \in \{0,...,T-1\}, \label{eq_constraint2} \\
 &&& \text{unknown } \vec{a}_{\tt J} (t), \forall t \in \{0,...,T-1\}, \label{eq_constraint3}
 \end{alignat}
\end{subequations}
where $\vec{A}$ in \eqref{eq_obj_func} can be expressed as
\begin{align} \label{eq_action_matrix}
 \vec{A} = [\vec{a}_{{\tt UD}_1}, \cdots, \vec{a}_{{\tt UD}_n}, \cdots, \vec{a}_{\mathtt{iUD}}].  
\end{align}
\section{The POMDP Formulated and the DQN Solution Proposed}
In modern wireless networks, decision-making problems are generally modeled using the MDP \cite{bozkus2023link}. A variant, the POMDP, is used when the agent has incomplete information about its environment. To be able to apply reinforcement learning (RL) algorithms to the channel allocation optimization problem, the problem, including the agent, the action space $\mathcal{A}$, the state space $\mathcal{S}$, the instantaneous reward function $r$, the transition probability $\mathcal{P}$, and the policy $\pi$ must be converted to the POMDP framework. 

At time $t$, the agent observes a state $s_t \in \mathcal{S}$ and takes action $a_t \in \mathcal{A}$, resulting in a new state $s_{t+1}$ with a transition probability $\mathcal{P}$ and an instantaneous reward $r_{t+1}$. The interaction continues until the agent learns the optimal policy $\pi^*$. The agent's policy $\pi$ maps states to actions $\mathcal{S} \leftarrow \mathcal{A}$, i.e., $a_t = \pi(s_t)$. 
The long-term reward is defined as the expected accumulated discounted instantaneous reward and given by $\mathcal{R} = \mathbb{E}[\sum_{t=1}^{T} \gamma r_{t+1}(s_t, \pi(s_t))]$. The parameter $0 \leq \gamma \leq 1$ is the discount factor. The main goal is to obtain the optimal decision policy $\pi^*$ that maximizes the long-term reward, i.e., \mbox{$\pi^* = \max_\pi \mathcal{R}$}. 
\subsection{The Proposed POMDP to Address the Formulated Optimization Problem}

In our SCLARM problem, the iUD can only see the current channel state at each time slot. However, it can infer a distribution over the system state based on past actions and observations. The SCLARM problem is therefore compatible with the POMDP framework \cite{wang2018deep}. Next, we define the SCLARM problem's important parameters in terms of the POMDP framework. The agent is the iUD that uses our algorithm to maximize its objective function. 
$\pi^*$(s) serves to utilize the spare time slots while avoiding collisions and evading jamming. In the $t^{th}$ time slot, the iUD determines whether to transmit over the channel or wait depending on the observations it receives from the $\mathrm{AP}$. 
\subsubsection{Action Space}
A discrete action space $\mathcal{A} \dot= \{a_i: i = 1, 2\}$ represents the iUD's transmission decisions at each time slot. The transmission decision at time step $t$ is defined as $a_i \in \{\tt{dispatch}, \tt{hold} \}$, where $\mathtt{dispatch}$ implies packet transmission and $\mathtt{hold}$ means no transmission. 
\subsubsection{Channel Observation}
After the iUD's action $a_t$ has been taken, the $\mathrm{AP}$ broadcasts its feedback $\mathrm{ACK}_{t+1}$, which represents the transmission status, over a control channel. The channel observation vector is an action-observation pair that is expressed as $\vec{c}_{t+1} = (a_{t}, \mathrm{ACK}_{t+1})$ and becomes a component of the system state.
\subsubsection{State space}
The state space \mbox{$\vec{\mathcal{S}} = [\vec{s}_1, \cdots, \vec{s}_t, \cdots, \vec{s}_T]$} contains all the possible states that the environment can transition into. The state at time step $t+1$, which is denoted by $\vec{s}_{t+1} \in \vec{\mathcal{S}}$, represents the real-time transmission status and transmission rate of the UDs and is expressed as
\begin{align}\label{eq:environmental_state}
\scalebox{0.91}{
$\vec{s}_{t+1} = \big[(\vec{c}_{t+1}^{\tt (UD_1)}, r_{t+1}^{\tt (UD_1)}), \cdots, 
(\vec{c}_{t+1}^{\tt (UD_n)} r_{t+1}^{\tt (UD_n)}), (\vec{c}_{t+1}^{\tt (iUD)}, r_{t+1}^{\tt (iUD)})\big].$
}
\end{align}

\subsubsection{Utility Functions}
We define the utility function of a legitimate $\mathrm{UD}_n$ $\in N \{n: 1, 2, \cdots, N\}$ as a function of the UD's achievable rate given by $\mathbb{U}_{t}^{\mathtt{UD}_{n}} =\nu^\mathtt{UD} \times R^{\mathtt{UD}_n}$ and $\mathbb{U}^{\tt iUD}_{t} = \nu^{\tt UD} \times R^{\tt iUD}$, where ${\nu^{\tt UD}}$ is the scaling factor that is given in Table \ref{table: scaling_factors}. More specifically, ${\nu^{\tt UD}}$ is used for iUD to distinguish among the networks' different scenarios. 
\begin{table}[htp!]
\centering
\caption{Scaling factor $\nu^{\tt UD}$ \& $\nu^{\tt net}$ and \textnormal{iUD} decision}
\label{table: scaling_factors}
\begin{tabular}{lllll}
\toprule
\textbf{Slot Status} & $a_t$ by iUD & iUD Decision & $\nu^{\tt UD}$ & $\nu^{\tt net}$ \\
\hline
Jammed & hold & Good (\tt G) & 4 & 5 \\
fUD Transmitting & hold & Good (\tt G) & 4 & 5 \\
Free & hold & Worst (\tt W) &1 & -10 \\
Jammed & dispatch & Worst (\tt W) & 1 & -10 \\
fUD Transmitting & dispatch & Bad (\tt B) & 3 & -5 \\
Free & dispatch & Excellent (\tt E) & 5 & 10 \\
\bottomrule
\end{tabular}
\end{table}
\subsubsection{Agent's Reward Function}
Since the iUD's objective is to maximize SCLAR, we associate the definition of reward with the utility values of all the legitimate UDs scaled by the scalar reward ${\nu^{\tt net}}$ and an observation from the $\mathrm{AP}$. The value of reward $r_{t+1}$ at different $O_{t+1}$ is influenced by $\mathbb{U}^{A_g}_{t}$ and $\sum_{n=1}^{N}\mathbb{U}_{t}^{\mathtt{UD}_{n}}$ such that $r_{t+1} = \nu^{\tt net}(\mathbb{U}_t^{\tt iUD} + \sum_{n=1}^{N} \mathbb{U}_t^{\tt {UD}_n})$ where $\nu^{\tt net}$ is the iUD's scaling factor set out in Table \ref{table: scaling_factors}.
\subsubsection{Transition Probability}
The state transition function \mbox{$\mathcal{P}: \mathcal{S} \times A \rightarrow \Pi(\mathcal{S})$} represents how the state changes \mbox{$s_t \rightarrow s_{t+1}$} with uncertainty. In our SCLARM formulation, \mbox{$\mathcal{P}$ depends} on the UD's actions and the jammer's status and is given by
\begin{align}\label{eq: transition_prob}
 \mathcal{P}(\vec{s}_{t+1} | \vec{s}_t, a_t) &= p(a_{\tt UD_1}|\vec{s}_t) \times p(a_{\tt UD_2}|\vec{s}_t) \times   \cdots \times p(a_{\tt iUD}|\vec{s}_t) \nonumber \\
 &\quad\times \cdots \times p(a_{\tt UD_N}|\vec{s}_t) \times p(a_{\tt J_m}|\vec{s}_t). 
\end{align}
In our case, determining \eqref{eq: transition_prob} presents a significant challenge when it comes to solving the optimization problem formulated in \eqref{eq_obj_func} because the iUD is not aware of network dynamics. However, as it is explained in \cite{chen2020overview}, historical trajectories generated by the iUD's interactions with the environment can be utilized to search for $\pi^*$. Thus, the iUD can learn a policy $\pi$ directly from its experience instead of learning $\mathcal{P}$. 
\subsection{The Proposed Approach to Solve the POMDP formulated}
Value-based methods are used to estimate the agent's value function. The value function is then utilized to implicitly obtain $\pi^*$. Two value functions are used in the literature to measure the expected accumulated discounted rewards of being in a certain state or taking a certain action in a state based on a policy $\pi$; the state-value function $V^{\pi}(s)$ and the action-value function $Q(s_t, a_t)$. More specifically, the function $V^{\pi}(s_t)$ indicates the value of being in a state $s_t$ when following a policy $\pi$. Mathematically speaking, the state-value function $V^{\pi}(s)$, the action-value function $Q(s_t, a_t)$, the optimal value function $V^*(s)$, and the optimal action-value function $Q^*(s_t, a_t)$ can be expressed as
$
    V^{\pi}(s)\! =\! \mathbb{E}\big[\sum_{t=0}^{\infty}\!\gamma r_{t+1}(s_t, a_t, s_{t+1})|a_t \!\sim\! \pi(.|s_t), s_0\!=\!s\big],
    Q^{\pi}(s,a) =  \mathbb{E}\big[\sum_{t=0}^{\infty}\gamma r_{t+1}(s_t, a_t, s_{t+1})|a_t \sim \pi(.|s_t), s_0=s, a_0=a\big],
    V^*(s) = \max_{a_t}[r_{t+1}(s_t, a_t) + \gamma \mathbb{E}_{\pi}V^*(s_{t+1})],\, \text{and}\,
    Q^*(s,a) = r_{t+1}(s_t, a_t) + \gamma \mathbb{E}[\max_{a_{t+1}}Q^*(s_{t+1}, a_{t+1})]
$, respectively.

In general, the RL agent's goal is to obtain the optimal policy $\pi^*(s) = \argmax_{a} Q^{*}(s, a)$. Various methodologies can be employed to approximate $\pi^*(s)$, each with its unique approach. Value-based methods, for instance, determine $\pi^*(s)$ by updating the $Q(s, a)$ until convergence is reached where $Q(s, a) \approx Q^*(s, a)$. Subsequently, the $\pi^*(s)$ is then derived from $Q^*(s, a)$. A prevalent value-based method used in wireless communication is Q-learning \cite{morales2020grokking}. We elaborate on Q-learning and its extensions below.  
\subsubsection{Q-Learning}
Q-learning is one of the algorithms most widely used to solve problems modeled using MDP. It is a model-free online method that learns by direct interaction with the environment. It updates estimates using observed rewards and transitions at each iteration and focuses on evaluating the quality of actions in each state to gain comprehensive experience. Q-learning obtains the optimal values of the Q-function using the following update rule that is based on the Bellman equation
\begin{align}\label{eq: q_learning_update_rule}
    Q(s_t, a_t) &\leftarrow Q(s_t, a_t) + \alpha[r_{t+1}(s_t, a_t) \nonumber \\
    &\qquad + \gamma\max_{a_{t+1}}Q(s_{t+1},a_{t+1}) - Q(s_t, a_t)],
\end{align}
where $\alpha$ is the learning rate, $\gamma$ is the discount factor, $r_{t+1}(s_t, a_t)$ is the reward at time step $t+1$ for action $a_t$ in state $s_t$, and $s_{t+1}$ is the subsequent state after action $a_t$ is taken in state $s_t$. This iterative process gradually produces a Q-table containing optimal Q-values that facilitate effective decision-making. However, traditional Q-learning suffers from slow convergence and has a limited capacity to handle high-dimensional and dynamic state and action spaces \cite{yu2023user}. These limitations are overcome by deep neural networks to approximate $Q(s, a)$, as discussed below.  
\subsubsection{Deep Q Network}
A deep Q network (DQN) algorithm integrates the favorable characteristics of Q-learning and deep learning techniques \cite{sivaranjani2023artificial}. It replaces the Q-table in Q-learning with a deep neural network (DNN) that approximates $Q(s, a; \vec{\theta})$ by adjusting $\vec{\theta}$, which represents the neural network parameters, through training. The action-value function for a given $\vec{\theta}$ can be computed approximately for unseen state-action pairs $(s_t, a_t)$. The DQN learns $\vec{\theta}$ instead of the $|\mathcal{S}| \times |\mathcal{A}|$ matrix of Q-values. The neural network tries to find $\vec{\theta}$ that approximates $ V^*(s)$ and $ Q^*(s,a)$ in every state-action pair $(s,a) \in \mathcal{S} \times \mathcal{A}$. This process is referred to as the DQN training cycle, which is elaborated next. 
\subsubsection{The DQN Training Cycle}
 At time step $t$, the state of the environment is $s_t$, and the ResDNN's parameters are $\vec{\theta}$. The agent takes an action $a_t = \argmax_a Q(s_t, a_t; \vec{\theta})$ for which the DQN gives an output of $Q(s_t, a_t;\vec{\theta})$ for all actions in $\mathcal{A}$. With reward $r_{t+1}$ and next state $s_{t+1}$, the tuple $(s_t, a_t, r_{t+1}, s_{t+1})$ forms an experience tuple that is utilized by the DQN for training. To find the optimal $\vec{\theta}$, the DQN minimizes the prediction errors of $Q(s_t, a_t; \vec{\theta})$ using the loss function \cite{morales2020grokking} given by 
\begin{align}\label{eq:dqn_loss}
\mathcal{L}_{\tt MSE}(\vec{\theta}) 
&= \big(y^{\tt True}_{r_{t+1},s_{t+1}} - Q(s_t, a_t ; \vec{\theta})\big)^2,
\end{align}
where $Q(s_t, a_t;\vec{\theta})$ is the estimated Q-value and $y^{\tt True}_{r_{t+1},s_{t+1}}$ is the DQN's target output, which can be calculated using the Bellman equation 
 $y^{\tt True}_{r_{t+1},s_{t+1}} = r_{t+1} + \gamma \max_{a_{t+1}} Q(s_{t+1}, a_{t+1}; \vec{\theta}).$ 
The DQN is trained by repeatedly updating $\vec{\theta}$ using a semi-gradient algorithm given by
$\vec{\theta}_{k+1} \leftarrow \vec{\theta}_k - \alpha_k \nabla_{\vec{\theta}} \mathcal{L}_{\tt MSE}(\vec{\theta}_k)$,
where $\alpha_k$ is the step size at iteration $k$ and $\nabla_{\vec{\theta}} \mathcal{L}_{\tt MSE}(\vec{\theta}_k)$ is the gradient of the loss function with respect to $\vec{\theta}$ at iteration $k$.

The DQN algorithm uses a ``quasi-static target network'' technique to maintain two networks: a prediction network $Q(s_t, a_t, \vec{\theta}^{\tt pred})$ and a target network $Q(s_t, a_t; \vec{\theta}^{\tt target})$, where $\vec{\theta}^{\tt pred}$ and $\vec{\theta}^{\tt target}$ refer to the weight parameters of their respective network. The target network helps stabilize the learning process by reducing the variance and correlation of the target values. The DQN trains the prediction network at each learning step to decrease the loss function in \eqref{eq:dqn_loss}.

The target network is typically a copy of the original DQN with $\vec{\theta}^{\tt target}$ replaced by $\vec{\theta}^{\tt pred}$. Instead of immediately following $\vec{\theta}^{\tt target}$ at each training iteration, the target network updates the values of $\vec{\theta}^{target}$ using a soft update rule \cite{kobayashi2021t}, which is expressed as
\begin{align}\label{eq: update_theta_target}
    \vec{\theta}^{\tt target} \leftarrow (1-\tau)\vec{\theta}^{\tt target} + (\tau)\vec{\theta}^{\tt pred},
\end{align}
where $\tau$ is a hyper-parameter that controls the frequency of target updates. This technique results in the target network being updated gradually and smoothly after a few iterations. Thus, the loss function \eqref{eq:dqn_loss} and update equation lead to 
\begin{align}\label{eq:loss_with_e_replay}
    \mathcal{L}_{\tt MSE_{B}}(\vec{\theta}) = \frac{1}{\tt N_{\tt BE}}\sum_{e \in \tt EB}{\big(y^{\tt True}_{r_{t+1},s_{t+1}}-q(s_t,a_t;\vec{\theta}^{\tt target})}\big)^2,
\end{align}
where $
 y^{\tt True}_{r_{t+1},s_{t+1}} = r_{t+1} + \gamma \max_{a_{t+1}} q (s_{t+1} , a_{t+1} ; \vec{\theta}^{\tt pred}) 
$
and
\begin{align}\label{eq: theta_update_final}
\vec{\theta}^{\tt pred}_{k+1} \leftarrow \vec{\theta}^{\tt pred}_k - \alpha_k \nabla_{\vec{\theta}} \mathcal{L}_{\tt MSE_{\tt B}}(\vec{\theta}^{\tt pred}_k; \vec{\theta}^{\tt target}_k).
\end{align}
In this work, to stabilize the loss function in \eqref{eq:loss_with_e_replay} we employ a ResDNN-based function approximation. The pseudocode followed by the iUD is given in Algorithm \ref{my_algo}. 
\begin{algorithm}[t]
\caption{DRL-based iUD channel access for varying frame sizes }\label{my_algo}
 {\bf Input: } $\tt {N_{UDs}}, p, \mathcal{A} , S, \epsilon, \gamma, \tau, \tt EB, \tt N_{BE}, \vec{\theta}^{pred}, \vec{\theta}^{target}$\\
{\bf Initialize: } $s_0 \in \mathcal{S}, a_0 \in \mathcal{A}$, set $s_{t+1}= 0, r_{t+1}=0$ \\
\nl $\tt env \leftarrow \tt Env(p, \alpha , s_0, \epsilon, \epsilon_{min}, \epsilon_{decay}, \gamma, \tau, N_{UDs})$ \\
\For {$X_f: \{f= 1, 2, \cdots, F\}$}{
 \For{$t \in X_f= \{0,1,2,\cdots\, T\}$} {
  Set $\lambda=1, \tt PLE=1, \Lambda=1$, \\
         $NC \sim \mathcal{U}(2\,\tt dBm,\,5\,\tt dBm)$\\
          $[P_{1}, \cdots, P_{N}, P_{\tt J}, P_{\tt {iUD}}] \sim \mathcal{U}(20\,\tt dBm,\,25\,\tt dBm)$ \\
          Compute $[h_{n}, g_{m}, h_{\tt {iUD}}]$ using $\lambda$ \& $\Lambda$ $\forall n\in N,\, \forall m\in M$. \\ 
      
          Obtain $\Gamma_n$ via \eqref{mf_n} $\forall n\in N$\\
          Calculate $C_{n}$ via \eqref{eq: physical_rate_frame} $\forall n\in N$\\
          Observe $s_0$ \& update $a_t \leftarrow \epsilon$-greedy([$s_0$]) \\ 
          Get $r_{t+1}, \mathrm{ACK}^{i}_{t+1}, s_{t+1}$ from $\tt env.step$ ([$a_t$])\\
          Deposit $(s_t, a_t, r_{t+1}, s_{t+1})$ to $\tt EB$\\
          Sample $\tt{E^{N_{BE}}}$ experiences  [$s_t^{\tt N_{BE}}$, $a_t^{\tt N_{BE}}$, $r_{t+1}^{\tt N_{BE}}$, $s_{t+1}^{\tt N_{BE}}]$ \\
         
          \For {each $e \in \tt{E^{N_{BE}}}$ = $(s_t, a_t, r_{t+1}, s_{t+1})$ }
             {
              Compute $Q(s_t, a_t; \theta^{\tt pred})$  \\
              Compute $Q (s_{t+1}, a_{t+1};\theta^{\tt target})$ \\
             
              Perform Gradient Descent to update $\vec{\theta}^{\tt pred}$\\ 
              Update $\vec{\theta}^{\tt pred}$ using \eqref{eq: theta_update_final} \\
             \If{$t\%\tau==0$}
                {
                 Update $\vec{\theta}^{\tt target}$ using \eqref{eq: update_theta_target}
                }
             }
          Optimal $\theta^{\tt pred} \rightarrow \pi^*(s)$
         \\
          Update $s_{t} \leftarrow s_{t+1}$\\
          Choose $a^*_t \leftarrow \pi^*(s)$
 }
 } 
\end{algorithm}
\section{Numerical Results and Discussion}\label{section: simulation_results}
In this section, we evaluate the proposed solution's performance. First, we examine the DRL agent's learning curve and training loss. Subsequently, we demonstrate iUD's performance at maximizing the network's SCLAR. We used Python for all our experimental simulations. The RL framework was simulated using Keras and TensorFlow.
\begin{figure}[t]
\centering
\includegraphics[width=.85\linewidth]{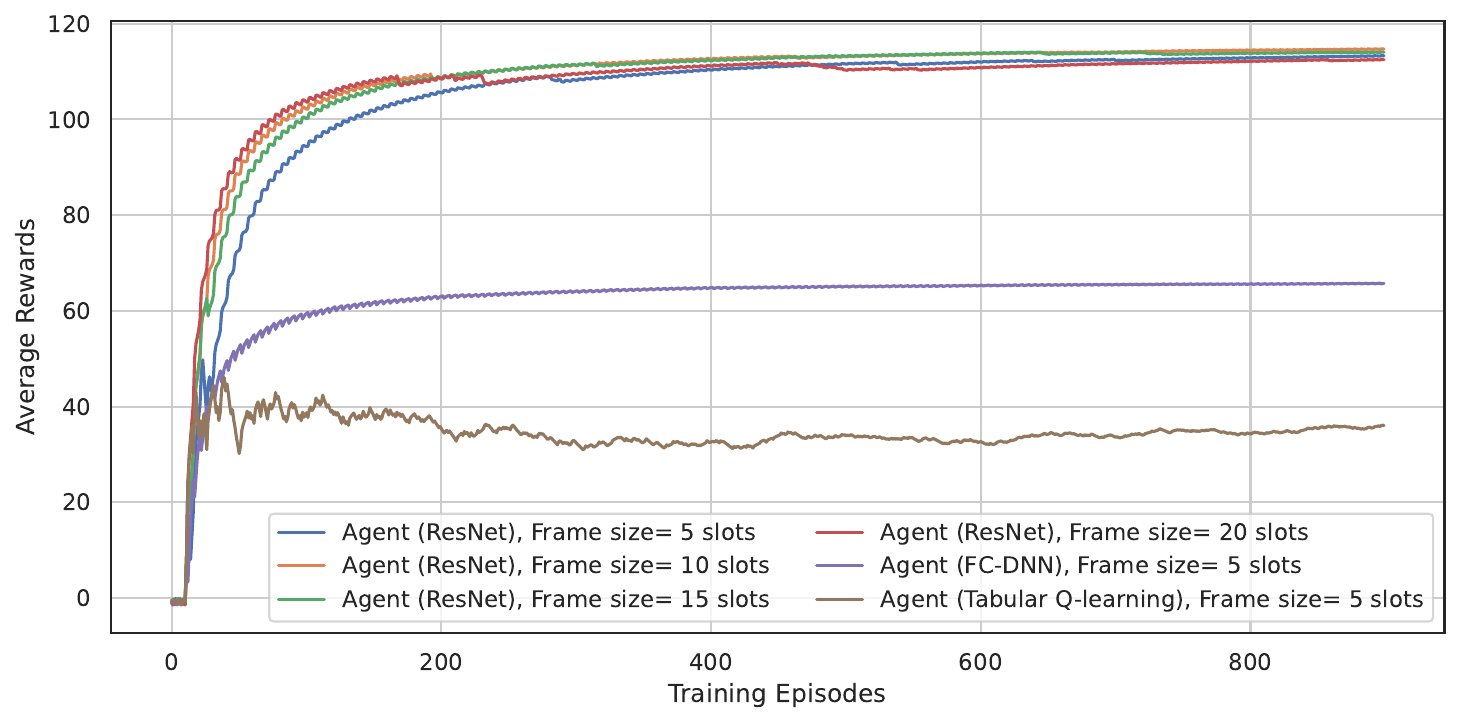}
\caption{The convergence curves of the learning agents.}
\label{fig:learning_curve}
\end{figure}

Fig.~\ref{fig:learning_curve} illustrates the convergence of agents using ResNet, FC-DNN, and tabular Q-learning for $Q(s_t,a_t)$ estimation. The ResNet-based agent converges at high average rewards regardless of frame size variations due to its superior capacity to learn complex patterns and generalize across states. In contrast, the FC-DNN-based agent settles at sub-optimal rewards since it lacks hierarchical feature learning, resulting in sub-optimal policies. Finally, the tabular Q-learning-based agent fails to converge due to the curse of dimensionality in large state spaces.

\begin{figure}[t]
\centering
\includegraphics[width=0.85\linewidth]{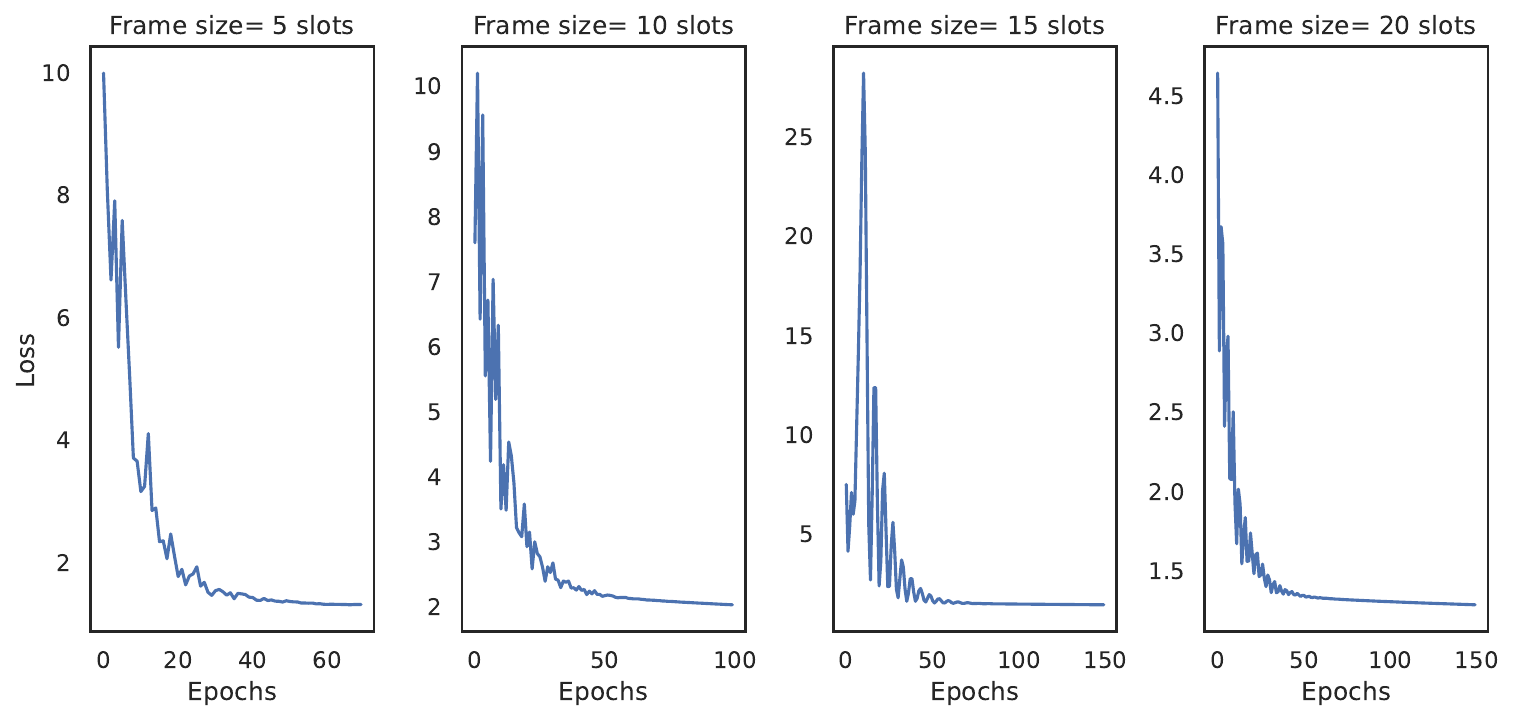}
\caption{ResDNN loss under different frame settings.}
\label{fig:epoch_loss}
\end{figure}

Fig.~\ref{fig:epoch_loss} shows the neural network's epoch loss against the number of training epochs. For all the frame sizes considered, the ResDNN-based DRL agent successfully reduces the loss in each successive epoch, which indicates improvement in its prediction capability and performance for all training episodes.
\begin{figure}[t]
        \centering
\includegraphics[width=.85\linewidth]{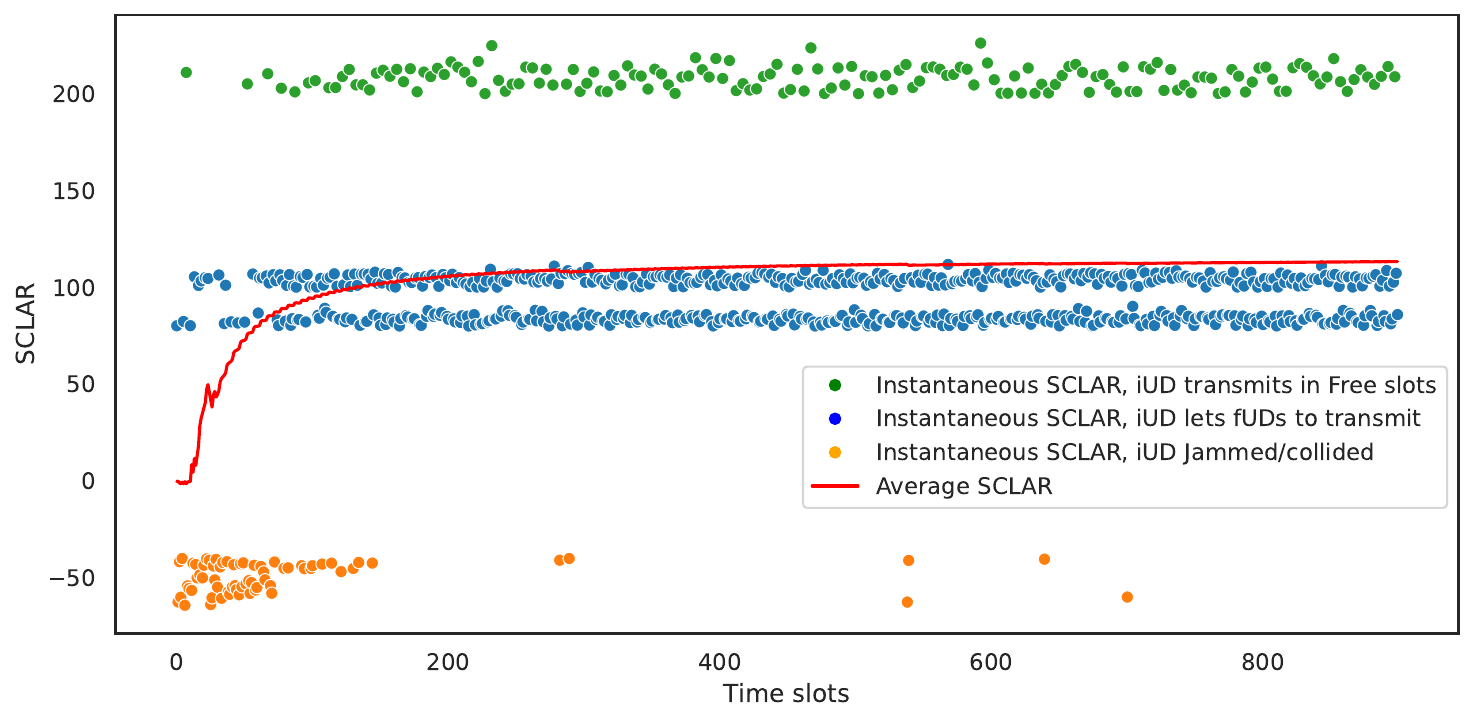}\label{fig:sclar} 
        \caption{SCLAR plot of the considered network.}
        \label{fig:CLAR_SCLAR}
\end{figure}
In Fig.~\ref{fig:CLAR_SCLAR}, we analyze the iUD's performance in terms of improving network SCLAR considering the frame size of 5 time slots. In the initial time slots, the iUD's packets frequently collide with those of the coexisting fUDs and are jammed by the jammer, resulting in the lowest SCLAR values. However, after a few time slots, the iUD can maximize the network's SCLAR and channel utilization by learning and implementing an optimal transmission strategy featuring negligible collisions and improved jamming robustness.
\section{Conclusion}\label{section: conclusions}
This paper presents a DRL-based iUD in a quasi-static network that learns a transmission strategy to maximize the network’s SCLAR. The proposed approach reduces collisions with legitimate UDs and avoids jamming. Simulation results show that the iUD learns to coexist with other UDs under changing channel conditions. After training, the iUD is able to select time slots for data packet transmission and maximize the network's SCLAR. To validate the proposed algorithm we compare our results with tabular Q-learning and a fully connected deep neural network approach. 
\bibliographystyle{IEEEtran}
\bibliography{main}

\begin{thebibliography}{10}
\providecommand{\url}[1]{#1}
\csname url@samestyle\endcsname
\providecommand{\newblock}{\relax}
\providecommand{\bibinfo}[2]{#2}
\providecommand{\BIBentrySTDinterwordspacing}{\spaceskip=0pt\relax}
\providecommand{\BIBentryALTinterwordstretchfactor}{4}
\providecommand{\BIBentryALTinterwordspacing}{\spaceskip=\fontdimen2\font plus
\BIBentryALTinterwordstretchfactor\fontdimen3\font minus
  \fontdimen4\font\relax}
\providecommand{\BIBforeignlanguage}[2]{{%
\expandafter\ifx\csname l@#1\endcsname\relax
\typeout{** WARNING: IEEEtran.bst: No hyphenation pattern has been}%
\typeout{** loaded for the language `#1'. Using the pattern for}%
\typeout{** the default language instead.}%
\else
\language=\csname l@#1\endcsname
\fi
#2}}
\providecommand{\BIBdecl}{\relax}
\BIBdecl

\bibitem{khadr2022jamming}
M.~H. Khadr, H.~B. Salameh, M.~Ayyash, H.~Elgala, and S.~Almajali, ``Jamming
  resilient multi-channel transmission for cognitive radio {IoT}-based medical
  networks,'' \emph{Journal of Communications and Networks}, vol.~24, no.~6,
  pp. 666--678, 2022.

\bibitem{miuccio2022learning}
L.~Miuccio, S.~Riolo, S.~Samarakoon, D.~Panno, and M.~Bennis, ``Learning
  generalized wireless {MAC} communication protocols via abstraction,'' in
  \emph{IEEE GLOBECOM 2022}.\hskip 1em plus 0.5em minus 0.4em\relax IEEE, 2022,
  pp. 2322--2327.

\bibitem{da2020noma}
M.~V. da~Silva, R.~D. Souza, H.~Alves, and T.~Abr{\~a}o, ``A {NOMA}-based
  {Q}-learning random access method for machine type communications,''
  \emph{IEEE Wireless Commun. Lett.}, vol.~9, no.~10, pp. 1720--1724, 2020.

\bibitem{mennes2020multi}
R.~Mennes, F.~A. De~Figueiredo, and S.~Latr{\'e}, ``Multi-agent deep learning
  for multi-channel access in slotted wireless networks,'' \emph{IEEE Access},
  vol.~8, pp. 95\,032--95\,045, 2020.

\bibitem{xin2022deep}
J.~Xin, W.~Xu, Y.~Cai, T.~Wang, S.~Zhang, P.~Liu, Z.~Guo, and J.~Luo, ``Deep
  learning based {MAC} via joint channel access and rate adaptation,'' in
  \emph{2022 IEEE 95th VTC2022-Spring}.\hskip 1em plus 0.5em minus 0.4em\relax
  IEEE, 2022, pp. 1--7.

\bibitem{fihri2020machine}
W.~F. Fihri, H.~El~Ghazi, B.~Abou El~Majd, and F.~El~Bouanani, ``A machine
  learning approach for backoff manipulation attack detection in cognitive
  radio,'' \emph{IEEE Access}, vol.~8, pp. 227\,349--227\,359, 2020.

\bibitem{khairy2020constrained}
S.~Khairy, P.~Balaprakash, L.~X. Cai, and Y.~Cheng, ``Constrained deep
  reinforcement learning for energy sustainable {multi-UAV} based random access
  {IoT} networks with {NOMA},'' \emph{IEEE J. Sel. Areas Commun.}, vol.~39,
  no.~4, pp. 1101--1115, 2020.

\bibitem{yu2019deep}
Y.~Yu, T.~Wang, and S.~C. Liew, ``Deep-reinforcement learning multiple access
  for heterogeneous wireless networks,'' \emph{IEEE J. Sel. Areas Commun.},
  vol.~37, no.~6, pp. 1277--1290, 2019.

\bibitem{morales2020grokking}
M.~Morales, \emph{Grokking deep reinforcement learning}.\hskip 1em plus 0.5em
  minus 0.4em\relax Manning Publications, 2020.

\bibitem{bozkus2023link}
T.~Bozkus and U.~Mitra, ``Link analysis for solving multiple-access {MDPs} with
  large state spaces,'' \emph{IEEE Trans. Signal Process.}, vol.~71, pp.
  947--962, 2023.

\bibitem{wang2018deep}
S.~Wang, H.~Liu, P.~H. Gomes, and B.~Krishnamachari, ``Deep reinforcement
  learning for dynamic multichannel access in wireless networks,'' \emph{IEEE
  Trans. Cognit. Commun. Networking}, vol.~4, no.~2, pp. 257--265, 2018.

\bibitem{chen2020overview}
S.~Chen and Y.~Li, ``An overview of robust reinforcement learning,'' in
  \emph{2020 IEEE International Conference on Networking, Sensing and Control
  (ICNSC)}.\hskip 1em plus 0.5em minus 0.4em\relax IEEE, 2020, pp. 1--6.

\bibitem{yu2023user}
W.~Yu, T.~J. Chua, and J.~Zhao, ``User-centric heterogeneous-action deep
  reinforcement learning for virtual reality in the metaverse over wireless
  networks,'' \emph{IEEE Trans. Wireless Commun.}, 2023.

\bibitem{sivaranjani2023artificial}
A.~Sivaranjani and B.~Vinod, ``Artificial potential field incorporated deep
  {Q}-network algorithm for mobile robot path prediction.'' \emph{Intelligent
  Automation \& Soft Computing}, vol.~35, no.~1, 2023.

\bibitem{kobayashi2021t}
T.~Kobayashi and W.~E.~L. Ilboudo, ``T-soft update of target network for deep
  reinforcement learning,'' \emph{Neural Networks}, vol. 136, pp. 63--71, 2021.

\end{thebibliography}
\end{document}